\documentclass[11pt,amsmath,amssymb,noeprint]{revtex4-1}

\usepackage{amsfonts}
\usepackage{amsmath}
\usepackage{amssymb}
\usepackage{array}
\usepackage{bbding}
\usepackage{bm}
\usepackage{braket}
\usepackage{breqn}

\usepackage{dcolumn}
\usepackage{diagbox}
\usepackage{endnotes}
\usepackage{epsfig}
\usepackage{epsf}
\usepackage{epstopdf}
\usepackage{extarrows}
\usepackage{graphicx}
\usepackage{harpoon}
\usepackage{hyperref}
\usepackage{indentfirst}
\usepackage{makecell}
\usepackage{mathrsfs}
\usepackage{mathtools}
\usepackage{multirow}
\usepackage{pifont}
\usepackage{xcolor}

\usepackage{cleveref}

\newcommand{\RM}[1]{\textrm{\uppercase\expandafter{\romannumeral#1}}}

\begin{document}

\title{Critical properties of bound states with one-boson-exchange potential}

\author{
Lin-Qing Song and Hai-Qing Zhou\protect\footnotemark[1]\protect\footnotetext[1]{E-mail: zhouhq@seu.edu.cn} \\
School of Physics, Southeast University, Nanjing 211189, China
}

\date{\today}

\begin{abstract}
In this study, we discuss some general critical properties of bound states with one-boson-exchange potential. For simplicity, we first take a system with two identical scalar particles as an example. The interaction between these two scalar particles is described by the exchange of another massive scalar meson under the instantaneous approximation, which results in the Yukawa potential. A highly accurate numerical method is used to determine the critical mass value of the system. The resulting critical mass for the ground state is consistent with those reported in the literature, agreeing to about 35 significant figures.  The highly accurate results for the $l=1$ case are also presented, which are significantly more precise than those previously reported in the literature. Furthermore, we extend the discussion to physical hadronic molecular states, where form factors are introduced in the interaction to describe the structure of hadrons. Our numerical results show that although the binding energies of the hadronic molecular states depend on the cutoff in the form factors, the number of hadronic molecular states is almost independent of the cutoffs across a very wide physical region. This indicates a strong and important property: for physically small couplings, the number of hadronic molecular states is almost solely determined by the coupling constants and the masses of the exchange particles.  This highly accurate numerical method can also be straightforwardly applied to higher $l$ cases or other systems.
\end{abstract}

\maketitle

\section{Introduction}

Hadronic molecular states have attracted considerable theoretical interest in hadronic physics since the experimental discovery of exotic hadron states, such as $X(3872)$, $Z_c(3900)$, $Z_b(10610)$, $Z_b(10650)$, as well as the $P_c$ and $P_{cs}$ states~\cite{Belle:2003nnu,BESIII:2013ris,Belle:2011aa,LHCb:2015yax,LHCb:2019kea,LHCb:2020jpq}.
Theoretically, these states are widely interpreted as hadronic molecular states. This picture is explored using various methods, most notably the Schr\"{o}dinger equation with one-boson-exchange (OBE) potential~\cite{Liu:2007bf,Yang:2011wz,Lee:2011rka,Zhao:2013ffn,Chen:2015loa,Chen:2017vai}, the unitary approach~\cite{Aceti:2014uea,Xiao:2013yca,Xiao:2019gjd,Feijoo:2021ppq,Duan:2022upr,Pan:2023hrk} and the Bethe-Salpeter equation~\cite{Sun:2011uh,He:2014nya,He:2015mja,Wang:2019ehs,Xu:2020gjl,Zhu:2021lhd,Ke:2022ocs,Liu:2023hrz}, where the interactions are based on the effective Lagrangian. In particular, owing to its intuitive physical insight, the potential model has been progressively refined and is frequently employed to describe hadronic molecular states, including the deuteron, based on the Yukawa potential. The Yukawa potential was proposed to explain the strong interaction between nucleons via the exchange of a massive particle~\cite{Yukawa:1935xg}. It has also been widely used in various branches of physics, such as plasmas, nuclear physics, astrophysics, solid-state physics
~\cite{debye1923theorie,PhysRevA.27.418,PhysRev.125.1131,PhysRev.134.A1235,Khrapak:2003kjw,DeLeo:2004vy,Loeb:2010gj,Chan:2013yza,PhysRev.178.1337,FERRAZ1984627}.

In the context of potentials for hadronic molecular states, form factors are used to describe the interaction between two non-point-like particles, which leads to a potential beyond the Yukawa form. These potentials are usually used to study the binding energies of the systems. The comparison of theoretical results with experimental data is used to fix the parameters in the form factors. The existence of these parameters in the form factors makes the situation more physical, but also makes the situation complex and obscures certain properties.

On the other hand, the study of the Schr\"{o}dinger equation with the Yukawa potential shows some interesting and beautiful global properties. For a system with two non-relativistic particles, each with mass $2\mu$, whose interaction is described by the Yukawa potential:
\begin{align}
V(r)=-\frac{\alpha}{r}e^{-m_{ex}r},
\end{align}
where $\alpha$ is the coupling constant and $m_{ex}$ represents the mass of the exchanged particle, the binding energies of such a system can be described by the Schr\"{o}dinger equation. When $m_{ex}=0$, the Yukawa potential reduces to the Coulomb potential, and there are an infinite number of bound states, as shown in textbooks. When $m_{ex}\neq 0$, the situation is fundamentally different. For any finite, positive $m_{ex}$, however, the number of bound states becomes finite. Moreover, there exists a critical mass beyond which no bound state can form. This phenomenon has been extensively investigated in previous studies~\cite{Xu:2021ret,Diaz_1991,PhysRevA.50.228,Li:2006as,Edwards:2017ndv,Napsuciale:2020ehf,Napsuciale:2021qtw} and the existence of an $S$-wave ground bound state is found to depend on the relationship among the coupling constant $\alpha$, the reduced mass $\mu$, and the exchange mass $m_{ex}$, which is given by the critical condition:
\begin{eqnarray}
	\frac{\mu}{\alpha m_{ex}} = 1.1906122105(5).
\label{equation:critical-screen-relation}
\end{eqnarray}
Furthermore, in Refs.~\cite{Napsuciale:2021qtw,Diaz_1991,delValle:2018eew}, the critical values of $m_{ex}$ for various angular momentum states were systematically calculated. These studies also provide the critical masses corresponding to the existence of the ground state, first excited state, second excited state, and higher excited states.

Physically, the number of bound states is a global property of the system, which is not determined by any single binding energy. For physical hadronic molecular states, when the hadrons' structure is considered, a natural question is what happens to the critical phenomenon in these systems. In this study, we focus on the critical properties of hadronic molecular states, rather than solely the binding energies. These critical properties indicate the global nature of the system and can help us understand its internal properties. To simplify the discussion, we take a single channel as an example.

The remainder of the article is organized as follows. In Sec.~\ref{Sec:Formalism}, we present the theoretical framework, detailing the derivation of the Schr\"{o}dinger equation and the potentials from the Bethe-Salpeter equation, and describe the numerical method employed. In Sec.~\ref{Sec:Numerical-results}, we present our numerical results, including the critical mass $m_{ex}$ required for the existence of bound states, the critical relationship between $m_{ex}$ and $\alpha$ with and without form factors, and how quantities depend on the cutoff $\Lambda$. Additionally, we give the computed binding energies as functions of $\alpha$ and $\Lambda$. Finally, Sec.~\ref{Sec:Summary} provides a summary of our results.

\section{Basic Formalism}\label{Sec:Formalism}

\subsection{From Bethe-Salpeter equation to Schr\"{o}dinger equation}
For simplicity, first we consider a system of two identical scalar particles, each with mass $m$, whose interaction is mediated by the exchange of a third scalar particle with mass $m_{ex}$. In Quantum Field Theory (QFT), the two-body bound state of this system is described by the Bethe-Salpeter (BS) equation:
\begin{align}
\chi_{\text{BS}}(P,p) = \int \frac{d^4k}{(2\pi)^4}K_{\text{BS}}(p-k)G(P,k)\chi_{\text{BS}}(P,k),
\label{equation:4D-BS-equation-1}
\end{align}
where $\chi_{\text{BS}}$ is the BS vertex, $K_{\text{BS}}$ is the BS irreducible interaction kernel, and $G(P,k)$ is the product of the two scalar particles' propagators, which is expressed as
\begin{align}
G(P, k)&=\frac{i}{k_1^2-m^2+i \epsilon} \frac{i}{k_2^2-m^2+i \epsilon}=\frac{-1}{(k_1^2-m^2+i \epsilon)(k_2^2-m^2+i \epsilon)},
\end{align}
where
\begin{align}
k_1&=\frac{1}{2}P+k,~~~~ k_2=\frac{1}{2}P-k,
\end{align}
with $k_1,k_2$ the momenta of the corresponding particles.

Generally, there is no analytic solution for the BS equation, even when $m_{ex} = 0$. When the binding energy is small compared to $m$, the instantaneous approximation is suitable, which means:
\begin{align}
K_{\text{BS}}(p-k) &\approx K_{\text{S}}(\boldsymbol{p}-\boldsymbol{k}).
\label{equation:K-instantaneous-approximation}
\end{align}

In the center-of-mass frame where $P\equiv(M,\boldsymbol{0})$, the instantaneous approximation and the form of $G(P,k)$ imply that $\chi_{\text{BS}}(P,p)$ is independent of $p_0$, leading to the following equation:
\begin{eqnarray}
\chi_{\text{S}}(M,\boldsymbol{p}) = \int \frac{d^4k}{(2\pi)^4}K_{\text{S}}(\boldsymbol{p}-\boldsymbol{k})G(P,k)\chi_{\text{S}}(M,\boldsymbol{k})=\int \frac{d^3\boldsymbol{k}}{(2\pi)^4}K_{\text{S}}(\boldsymbol{p}-\boldsymbol{k})\bar{G}(M,\boldsymbol{k})\chi_{\text{S}}(M,\boldsymbol{k}),
\label{equation:3D-Salpeter-equation-1}
\end{eqnarray}
where
\begin{align}
\chi_{\text{S}}(M,\boldsymbol{p}) & \equiv  \chi_{\text{BS}}(P,p) \Big|_{\text{instantaneous approximation}}, \notag\\
\bar{G}(M,\boldsymbol{k}) &\equiv \int dk_0 G(M,k)=\frac{2i\pi}{\omega_k}\frac{1}{M^2-4\omega_k^2+i\epsilon},
\end{align}
with $\omega_k=\sqrt{m^2+\boldsymbol{k}^2}$. By defining
\begin{align}
\phi_{\text{S}}(M,\boldsymbol{k})\equiv  \bar{G}(M,\boldsymbol{k})\chi_{\text{S}}(M,\boldsymbol{k}),
\end{align}
Eq.~\eqref{equation:3D-Salpeter-equation-1} can be equivalently rewritten as
\begin{align}
\phi_{\text{S}}(M,\boldsymbol{p}) =  \bar{G}(M,\boldsymbol{p}) \int\frac{d^3\boldsymbol{k}}{(2\pi)^4} K_{\text{S}}(\boldsymbol{p}-\boldsymbol{k})\phi_{\text{S}}(M,\boldsymbol{k}),
\end{align}
which is the usual form of the Salpeter equation. Since $E_b\equiv M-2m$ and $\boldsymbol{p}$ are much smaller than $m$, we can expand $\bar{G}(M,\boldsymbol{k})$ in terms of these two small quantities to leading order, which yields the Schr\"{o}dinger equation in momentum space:
\begin{align}(E_b-\frac{\boldsymbol{p}^2}{m}+i\epsilon)\phi(P,\boldsymbol{p}) &= \frac{i\pi}{2m^2}\int\frac{d^3\boldsymbol{k}}{(2\pi)^4}K_{\text{S}}(\boldsymbol{p}-\boldsymbol{k}) \phi(P,\boldsymbol{k}) = \int \frac{d^3\boldsymbol{k}}{(2\pi)^3}\Big[\frac{i K_{\text{S}}(\boldsymbol{p}-\boldsymbol{k})}{4m^2 } \Big] \phi(P,\boldsymbol{k}).
\end{align}
The corresponding Schr\"{o}dinger equation in coordinate space is then given by:
\begin{align}
(E_b-\frac{\hat{\boldsymbol{p}}^2}{2\mu}+i\epsilon)\psi(\boldsymbol{r}) &=V(\boldsymbol{r}) \psi(\boldsymbol{r}),
\end{align}
where $\mu=\frac{1}{2}m$ and
\begin{align}
\psi(\boldsymbol{r}) &\equiv \int d^3{\boldsymbol{q}}e^{i\boldsymbol{q}\cdot\boldsymbol{r}} \phi(P,\boldsymbol{q}), \notag \\
V(\boldsymbol{r}) &\equiv \int \frac{d^3{\boldsymbol{q}}}{(2\pi)^3}e^{i\boldsymbol{q}\cdot\boldsymbol{r}} i\bar{K}_{\text{S}}(\boldsymbol{q}),
\end{align}
with
\begin{align}
\bar{K}_{\text{S}}(\boldsymbol{q})\equiv \frac{1}{4m^2}K_{\text{S}}(\boldsymbol{q}).
\end{align}

\textcolor{black}{In the present work, the Schr\"{o}dinger equation is derived from the BS equation by adopting the instantaneous approximation, which effectively reduces the relativistic two-body dynamics to a non-relativistic framework. This approximation neglects some relativistic effects and assumes that the interaction kernel is independent of the relative time. It may become inadequate when relativistic corrections and energy-dependent effects are sizable and are not fully accounted for in the present treatment. Nevertheless, the Schr\"{o}dinger equation provides a well-established and widely used framework for investigating bound-state systems. In practical applications, part of the relativistic effects can be effectively absorbed into model parameters, such as an effective coupling constant or the effective mass of the exchanged particle. Therefore, the present approach is expected to provide a reasonable description at the qualitative and semi-quantitative level.}

For point-like scalar particles, taking the interaction vertex for the three scalar fields as $\Gamma=-2igm$ gives the interaction kernel:
\begin{align}
\bar{K}_{\text{S}} \rightarrow \bar{K}_{0}(\boldsymbol{p}-\boldsymbol{k}) & =\frac{ig^2}{(\boldsymbol{p}-\boldsymbol{k})^2+m_{ex}^2-i\epsilon},
\end{align}
and the corresponding potential in the coordinate space is expressed as
\begin{align}
V \rightarrow V_{0}(\boldsymbol{r}) &=-\frac{\alpha }{r}e^{-m_{ex}r},
\end{align}
with $r\equiv |\boldsymbol{r}|,\alpha\equiv \frac{g^2}{4\pi}$. This is simply a pure Yukawa potential. When the particles are not point-like, form factors are usually introduced to describe their structure, and the corresponding correction in momentum space modifies the kernel as:
\textcolor{black}{
\begin{eqnarray}
K_{\text{S}} \rightarrow K_{1}(\boldsymbol{p}-\boldsymbol{k}) & = \bar{K}_{0}(\boldsymbol{p}-\boldsymbol{k})\Big[\frac{\Lambda^2}{(\boldsymbol{p}-\boldsymbol{k})^2+\Lambda^2}\Big]^i,
\end{eqnarray}
where $i=2$ or $i=4$ corresponds to introducing a monopole-type or dipole-type form factor at each vertex, respectively. Because the kernel involves two vertices, the choices $i=2$ and $i=4$ amount to monopole and dipole form-factor schemes, respectively. The corresponding potentials in coordinate space with monopole and dipole form factors are expressed, respectively, as
\begin{eqnarray}
V_{1}(r) &=&-\frac{\alpha\Lambda^4}{2r(m_{ex}^2-\Lambda^2)^2}\Big[2 e^{-m_{ex}r}-2 e^{-\Lambda r}+e^{-\Lambda r}\frac{(m_{ex}^2-\Lambda^2)r}{\Lambda}\Big],\nonumber\\
V_{2}(r)&=&-\frac{ \alpha \Lambda^8}{48 r(m_{ex}^2-\Lambda^2)^4}\Big[48\ e^{-m_{ex} r}-48e^{-\Lambda r}+e^{-\Lambda r}P(m_{ex},\Lambda,r)\Big]
\end{eqnarray}
where
\begin{eqnarray}
P(m_{ex},\Lambda,r)&=&\frac{1}{\Lambda^5}\Big[m_{ex}^6 r[3+r\Lambda(3+r\Lambda)]-3m_{ex}^4r\Lambda^2[5+r\Lambda(5+r\Lambda)] \nonumber\\
&&+3m_{ex}^2r\Lambda^4[15+r\Lambda(7+r\Lambda)]-\Lambda^5[r\Lambda(33+r\Lambda(9+r\Lambda))]\Big].
\end{eqnarray}}

The above discussion can also be extended to systems involving particles with other quantum numbers, where the situation remains similar. After separating the angular part, similar to the hydrogen case, the radial part of the equation in coordinate space is given as in standard textbooks:
\begin{eqnarray}
\Big[-\frac{1}{2\mu} \frac{\partial^2}{\partial r^2} + \frac{l(l+1)}{2\mu r^2} + V_i(r)\Big]\chi(r)=E_b\chi(r),
\label{eq:radial-Schrodinger-equation}
\end{eqnarray}
where $l$ is the angular momentum quantum number, $\chi(r)\equiv r R(r)$ and $\psi(\boldsymbol{r})\equiv R(r)Y_{lm}(\Omega_{\boldsymbol{r}})$.

\subsection{Numerical method}
Since the Schr\"{o}dinger equation with the above general potential cannot be solved exactly~\cite{Edwards:2017ndv}, we employ the numerical shooting method to solve the system with high precision. In the practical calculation, we adopt the approximation that the wave function $\chi(r)$ satisfies the boundary conditions:
\begin{eqnarray}
	\chi(\delta) = 0,~~\chi'(\delta) = 1,~~\chi(\Delta) = 0,
\end{eqnarray}
where $\delta$ and $\Delta$ are parameters that correspond to the physical limits approaching 0 and $\infty$, respectively.

In the shooting method, we first fix $E_b$ and $m_{ex}$, and use the conditions $\chi(\delta)=0$ and $\chi'(\delta) = 1$ to solve Eq.~(\ref{eq:radial-Schrodinger-equation}) in the region $r\in[\delta,\Delta]$, thereby obtaining the value of $\chi(E_b,m_{ex},\Delta)$. To determine the binding energy for a known $m_{ex}$, we solve the equation  $\chi(E_b,m_{ex},\Delta)=0$  for the root $E_b$. To determine the critical value of $m_{ex}$, we  solve the equation $\chi(0, m_{ex},\Delta)=0$ for the root $m_{ex}$. We employ a Mathematica script to perform this procedure, utilizing the NDSolve function for solving the differential equation and FindRoot for locating the final root.  High accuracy is ensured by setting WorkingPrecision=200 and PrecisionGoal=40.

In the practical calculation of the binding energy for fixed $m_{ex}$ and $\alpha$ with $\mu=1$, we set the parameters in the following range:
\begin{eqnarray}
	\delta = 10^{-55} \sim 10^{-15},~~\Delta = 10^3\sim 10^{4}.
\label{eq:numerical-approximation}
\end{eqnarray}
where units are neglected.

In the practical calculation for the critical value of $m_{ex}$ with $\mu=1$ and $\alpha=1$, we set the parameters in the following range:
\begin{eqnarray}
\delta = 10^{-55} \sim 10^{-15},~~\Delta = 10^5\sim 10^{55}.
\end{eqnarray}
It is important to note that a significantly larger $\Delta$ is required for the critical points of the state, as the binding energy is zero. Consequently, the final results are only considered acceptable when they become stable with respect to the chosen ranges of $\delta$ and $\Delta$.  \textcolor{black}{For example, Tab.~\ref{tab:convergence-behavior} lists the critical ground-state $m_{ex}$ as a function of the integration parameters $\delta$ and $\Delta$ (cf.~the exponents in columns two and three), which clearly demonstrates the stability of the results.
\renewcommand\tabcolsep{0.6cm}
\renewcommand{\arraystretch}{1.5}
\begin{table}[htbp]
\centering
\begin{tabular}{cccc}\hline
   potential & $\log_{10}\delta$ & $\log_{10}\Delta$  &critical ground-state $m_{ex}$ \\\hline
    $V_0(r)$ &  -25&  25& 1.1906124210606177053427765163846311\\
          &   -35 &35  &1.1906124210606177053427771063610463\\
          &   -45 &45  &1.1906124210606177053427771063610463\\
          &  -55 &55  &1.1906124210606177053427771063610463\\\hline
$V_1(r)$ &  -25&  25  &0.67694250611421318125252911459994898\\
            &  -35&  35 &0.67694250611421318125252931447993144\\
            &  -45&  45 &0.67694250611421318125252931447993146\\
            &   -55& 55 &0.67694250611421318125252931447993146\\\hline
$V_2(r)$ &  -25&  25  &0.58558680838334368160717244234344288\\
             &  -35&  35 &0.58558680838334368160717264884765732\\
            &  -45&  45 &0.58558680838334368160717264884765735\\
            &   -55& 55 &0.58558680838334368160717264884765735\\
\hline
\end{tabular}
\caption{Critical ground-state $m_{ex}$ as a function of $\log_{10}\delta$ and $\log_{10}\Delta$, where $\delta=10^{k_\delta}$ and $\Delta=10^{k_\Delta}$ with the integers $k_\delta,k_\Delta$ given in the table (natural units as elsewhere in this section), for each model potential.} \label{tab:convergence-behavior}
\end{table}
}

Using the numerical method outlined above, we compare our numerical results with the analytical solutions for the binding energies under the Coulomb potential case with parameters $m_{ex}=0$, $\alpha=1$, and $\mu=1$. \textcolor{black}{The binding energies calculated via the numerical shooting method are consistent with the analytical solutions for the Coulomb potential.} Additionally, the comparison shows that our numerical results are consistent with the analytical solutions to 35 significant digits. This high level of precision confirms the validity of our approximation and numerical method. For the case where $m_{ex}\neq 0$, we expect this method to yield a similar precision.

\section{Numerical Results} \label{Sec:Numerical-results}

\subsection{Results for the Yukawa potential $V_0(r)$}
For the Yukawa potential $V_0(r)$, the critical value of $m_{ex}$ has been studied in Refs.~\cite{Diaz_1991,Edwards:2017ndv,delValle:2018eew,Napsuciale:2020ehf,Napsuciale:2021qtw,Jiao_2021,PhysRevE.108.045301}. As a first test and comparison, we present our numerical results for the critical value of $m_{ex}$ for the ground state in Tab.~\ref{1-critical-mex-different-Ref-V0}, where we fix the coupling constant $\alpha = 1$ and the reduced mass $\mu = 1$. The comparison shows that our results are consistent with those reported in Refs.~\cite{delValle:2018eew,Jiao_2021,PhysRevE.108.045301} with an accuracy of about 30 significant digits, which means that the shooting method is effective for obtaining high-precision results.

The critical values of $m_{ex}$ for states with other quantum numbers $n$ and $l$ are presented in Tabs.~\ref{tab:critical-mex-n2},\ref{tab:critical-mex-n3},\ref{tab:critical-mex-n4},\ref{tab:critical-mex-n5}. Most of our results are consistent with those reported in Refs.~\cite{Diaz_1991,Napsuciale:2021qtw,delValle:2018eew}. However, a significant discrepancy is still observed in some cases. Given that the approach employed in this work demonstrates sufficient accuracy for subsequent calculations, we conclude that the results obtained using the shooting method are reliable.  The combination of these results also shows a global property: for a fixed $n$, the critical values of $m_{ex}$ in the $l=1$ case are always smaller than those in the $l=0$ case.

In Tabs.~\ref{tab:critical-mex-n2}--\ref{tab:critical-mex-n5}, we only present our numerical results to 35 significant digits. In practical calculations, one can increase parameters such as $\text{WorkingPrecision}$ and $\text{PrecisionGoal}$ to achieve higher accuracy. Furthermore, results for other states, such as $l=2$, can also be easily obtained but are not listed here. The complete results demonstrate that the shooting method can yield highly accurate critical values of $m_{ex}$ for high angular momentum states ($l$). These high-precision values have not been previously reported in the literature.

\renewcommand\tabcolsep{0.2cm}
\renewcommand{\arraystretch}{1.0}
\begin{table}[htbp]
	\centering
		\begin{tabular}{c|c|l}\hline
			work &method &value of $m_{ex}$\\\hline
			Ref.~\cite{Diaz_1991}&matrix propagation &1.190612421060618\\
			Ref.~\cite{Edwards:2017ndv}& fifth-order perturbative calculation&1.1906122105(5)\\
			Ref.~\cite{Napsuciale:2020ehf,Napsuciale:2021qtw}&hidden
			supersymmetry and systematic expansion&1.1906124207(2)\\
				Ref.~\cite{delValle:2018eew}&the Lagrange Mesh
				Method (LMM)&1.190612\\
			Ref.~\cite{delValle:2018eew}&perturbative calculation and Pad\'{e} approximations (PT)&1.19061242106061770\\
			Ref.~\cite{Jiao_2021}&the generalized pseudospectral (GPS) method&1.190612421060617705342777106362\\
			Ref.~\cite{PhysRevE.108.045301}& coupled first-order
			differential equations&1.19061242106061770534277710636105\\
			this work&shooting method &1.1906124210606177053427771063610463\\
			\hline
\end{tabular}
\caption{Comparison of the critical value of $m_{ex}$ for the ground state where the parameters are taken as $\mu=1,\alpha=1$. }
\label{1-critical-mex-different-Ref-V0}
\end{table}

\renewcommand\tabcolsep{0.07cm}
\renewcommand{\arraystretch}{1.3}
\begin{table}[htbp]
\centering
\begin{tabular}{c|l|l}\hline
$n=2$& $l=0$ & $l=1$ \\\hline
Ref.~\cite{Diaz_1991} & 0.310209282713937 & 0.22021680661 \\
Ref.~\cite{Napsuciale:2021qtw} & 0.3102092834 (2) & 0.220118 (1) \\
LMM in Ref.~\cite{delValle:2018eew} & 0.310209 & 0.220216806605 \\
PT in Ref.~\cite{delValle:2018eew} & 0.31020928271393593 & 0.220216275 \\
this work & 0.31020928271393693911011221295317871  & 0.22021680660657304040504146328957711 \\
\hline
\end{tabular}
\caption{Comparison of the critical values of $m_{ex}$ for the states with principal quantum number $n=2$ and angular momentum quantum number $l=0,1$ in $V_0(r)$ case, where the parameters are taken as $\mu=1,\alpha=1$.}
\label{tab:critical-mex-n2}
\end{table}

\begin{table}[htbp]
\centering

\begin{tabular}{c|l|l}\hline
$n=3$ & $l=0$ & $l=1$  \\\hline

Ref.~\cite{Diaz_1991} & 0.13945029406418 & 0.11271049836  \\
Ref.~\cite{Napsuciale:2021qtw} & 0.139450295 (1) & 0.11265 (1) \\
LMM in Ref.~\cite{delValle:2018eew} & 0.139450 & 0.112710498359 \\
PT in Ref.~\cite{delValle:2018eew} & 0.13945029406417801 & 0.1127086527 \\
this work &0.13945029406417801388235795488972252 &0.11271049835952494497397295215522491  \\
\hline
\end{tabular}
\caption{Comparison of the critical values of $m_{ex}$ for the states with principal quantum number $n=3$ and angular momentum quantum number $l=0,1$ in $V_0(r)$ case, where the parameters are taken as $\mu=1,\alpha=1$. }
\label{tab:critical-mex-n3}
\end{table}

\begin{table}[htbp]
\centering

\begin{tabular}{c|l|l}\hline
$n=4$ & $l=0$ & $l=1$  \\\hline
Ref.~\cite{Diaz_1991} & 0.078828110273172 & 0.06788537610 \\
Ref.~\cite{Napsuciale:2021qtw} & 0.0788281106 (1) & 0.067827 (1)  \\
LMM in Ref.~\cite{delValle:2018eew} & 0.078428 & 0.067885376100  \\
PT in Ref.~\cite{delValle:2018eew} & 0.0788281102731706 & -- \\
this work &0.078828110273171565170282204980085099& 0.067885376100579552788417968577926856  \\
\hline
\end{tabular}
\caption{Comparison of the critical values of $m_{ex}$ for the states with principal quantum number $n=4$ and angular momentum quantum number $l=0,1$ in $V_0(r)$ case, where the parameters are taken as $\mu=1,\alpha=1$. The notation "--" signifies the absence of calculation in the corresponding reference.}
\label{tab:critical-mex-n4}
\end{table}

\begin{table}[htbp]
\centering
\begin{tabular}{c|l|l}\hline
$n=5$ & $ l=0$ & $l=1$ \\\hline
Ref.~\cite{Diaz_1991} & 0.050583170560 & 0.045186248  \\
Ref.~\cite{Napsuciale:2021qtw} & 0.0505831707 (2) & 0.045155 (1)  \\
LMM in Ref.~\cite{delValle:2018eew} & 0.050583 & 0.045186248071  \\
PT in Ref.~\cite{delValle:2018eew} & 0.0505831703745 & --  \\
this work & 0.050583170374558799782284408667887461 & 0.045186248071624990093706122691149700   \\
\hline
\end{tabular}
\caption{Comparison of the critical values of $m_{ex}$ for the states with principal quantum number $n=5$ and angular momentum quantum number $l=0,1$ in $V_0(r)$ case, where the parameters are taken as $\mu=1,\alpha=1$. The notation "--" signifies the absence of calculation in the corresponding reference.}
\label{tab:critical-mex-n5}
\end{table}

\begin{figure}[htbp]
\centering
\includegraphics[bb=25 140 400 250, clip,scale=1.5]{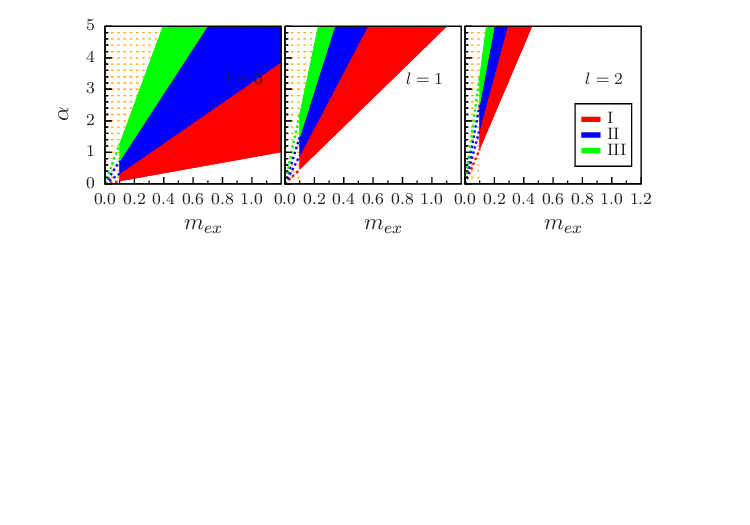}
\caption{Critical boundaries for the number of bound states as functions of $\alpha$ and $m_{ex}$ for specific angular momentum $l$ in the $V_0(r)$ case, where $\mu$ is fixed as $1$. The panels, ordered from left to right, illustrate the cases for $l=0$, $l=1$, and $l=2$. The red (\RM1), blue (\RM2), and green (\RM3) regions signify the existence of one, two, and three bound states, respectively.}
\label{Fig:bound-states-alpha-mex-V0-monopole}
\end{figure}

In Fig.~\ref{Fig:bound-states-alpha-mex-V0-monopole}, we show how the number of bound states evolves with respect to the parameters $m_{ex}$ and $\alpha$. The panels, ordered from left to right, illustrate the cases for $l=0$, $l=1$, and $l=2$. The red (\RM1), blue (\RM2), and green (\RM3) regions indicate the presence of one, two, and three bound states for the given $l$. The region marked with dots denotes an area that falls outside the scope of our discussion. Our calculations specifically consider the region where $m_{ex}\geq0.1$. The dashed lines show the critical-boundary curves obtained from our discrete numerical points, extrapolated toward smaller $m_{ex}$ in the $(\alpha,m_{ex})$ plane as justified below. \textcolor{black}{We extrapolate the critical boundaries toward the origin for two reasons. First, when $m_{ex}\rightarrow 0$, bound states exist for any sufficiently small positive $\alpha$, whereas for $\alpha=0$ there are none, so both $\alpha$ and $m_{ex}$ must approach zero along the critical line. Second, within the computed parameter range we observe an approximately linear relation between the critical coupling $\alpha$ and $m_{ex}$. Following this trend, the boundary curves are extended to smaller $m_{ex}$ and appear to converge visually to the origin as $\alpha\rightarrow 0$.} For the specific case of $l=0$ and $\alpha=1$, the critical value of $m_{ex}$ is approximately $1.19$, a result consistent with the data presented in Tab.~\ref{1-critical-mex-different-Ref-V0}. We observe that the slope of the boundary curve increases with the orbital angular momentum quantum number $l$. This indicates that a larger coupling constant is necessary to form a bound state as $l$ increases. Furthermore, the linear behavior of the region boundaries directly indicates that the critical values satisfy relations akin to Eq.~(\ref{equation:critical-screen-relation}). Such relations, in fact, can be determined through dimensional analysis.

\subsection{Results for potential $V_1(r)$}\label{Sec:n2}

The potential $V_1(r)$ is widely used in the study of molecular states using OBE models. Its critical parameters reflect the global properties of the system and are crucial for understanding the internal properties of the bound states. In this subsection, we discuss the related properties for this potential. For simplicity, we still use the scalar system as an example.

\renewcommand\tabcolsep{0.3cm}
\renewcommand{\arraystretch}{1.0}
\begin{table}[htbp]
\centering
\begin{tabular}{c|cc}\hline
\diagbox[height=0.3\baselineskip]{$n$}{$l$}& 0 & 1 \\\hline
 1 & 0.67694250611421318125252931447993146&--\\
 2& 0.20562081503354359818299315422443232&0.20695075065871335259276932355973134\\
 3& 0.10387929422723629079373251602398867&0.10384898223045325459958290388192965\\
 4& 0.062812729606088808273694095206222191&0.063137840555479296025152192163701117\\
 5& 0.042070819608385184596535181069213768&0.042454393767796674198021311139945282\\\hline
\end{tabular}
\caption{The critical values of $m_{ex}$ with $\mu=1$, $\alpha=1$, and $\Lambda=1$, categorized by quantum numbers $n$ and $l$ in the $V_1(r)$ case.}
\label{Tab:critical-mex-V1-monopole}
\end{table}

\begin{figure}[htbp]
\centering
\includegraphics[bb=25 10 400 250, clip,scale=1.5]{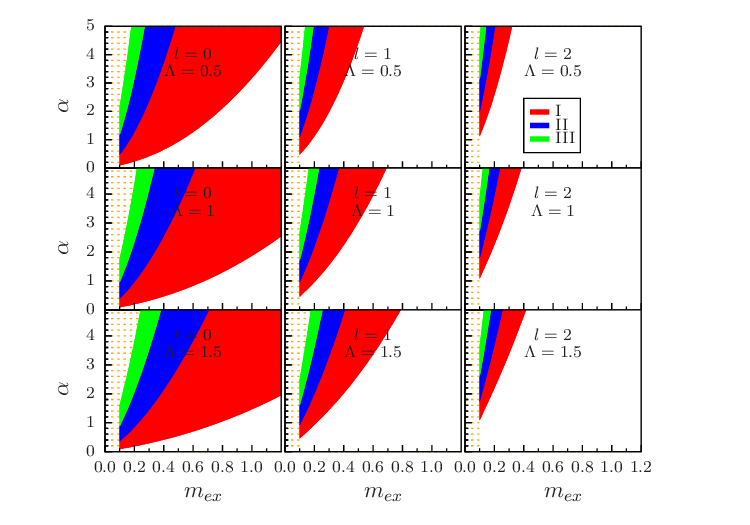}
\caption{Critical boundaries for the number of bound states as functions of $\alpha$ and $m_{ex}$ for specific angular momentum $l$ and cutoff $\Lambda$ in the $V_1(r)$ case, where $\mu$ is fixed as $1$. The panels, ordered from left to right, illustrate the cases for $l=0$, $l=1$, and $l=2$. The red (\RM1), blue (\RM2), and green (\RM3) regions signify the existence of one, two, and three bound states, respectively.}
\label{Fig:alpha-on-mex-V1-monopole}
\end{figure}

\begin{figure}[htbp]
\centering
\includegraphics[bb=25 140 400 250, clip,scale=1.5]{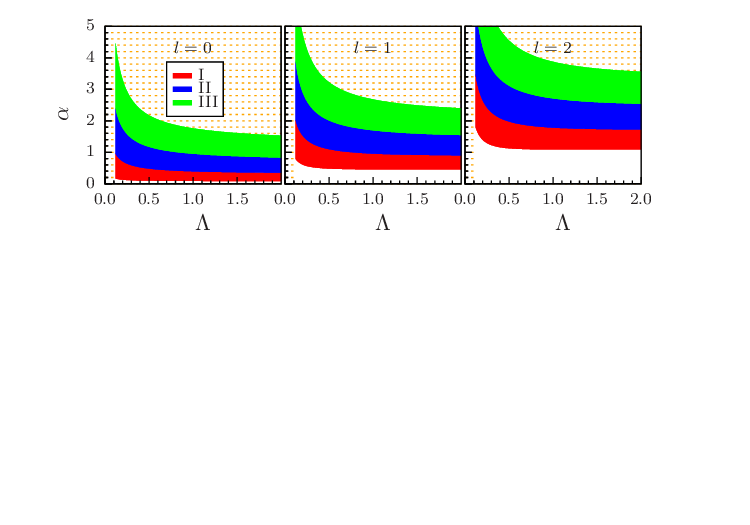}
\caption{Critical boundaries for the number of bound states as functions of $\alpha$ and $\Lambda$ for specific angular momentum $l$  in the $V_1(r)$ case, where $\mu$ is fixed as $1$ and $m_{ex}$ is fixed as $0.1$. The panels, ordered from left to right, illustrate the cases for $l=0$, $l=1$, and $l=2$. The red (\RM1), blue (\RM2), and green (\RM3) regions signify the existence of one, two, and three bound states, respectively.}
\label{Fig:alpha-on-Lambda-V1-monopole}
\end{figure}

Tab.~\ref{Tab:critical-mex-V1-monopole} provides the critical values of $m_{ex}$ for $\mu=1$, $\alpha=1$, and $\Lambda=1$, categorized by quantum numbers $n$ and~$l$. \textcolor{black}{Since there are no comparable results in the literature for the critical mass in the $V_1$ case, Tab.~\ref{Tab:critical-mex-V1-monopole} lists only our values.}  Fig.~\ref{Fig:alpha-on-mex-V1-monopole} visualizes how the number of bound states evolves with the parameters $m_{ex}$ and~$\alpha$ (with $\Lambda$ and $\mu$ fixed), using the same notations as in Fig.~\ref{Fig:bound-states-alpha-mex-V0-monopole}. Similarly, Fig.~\ref{Fig:alpha-on-Lambda-V1-monopole} shows how the number of bound states evolves with the parameters $\alpha$ and $\Lambda$ with $m_{ex}$ and $\mu$ fixed.

The results in Tab.~\ref{Tab:critical-mex-V1-monopole} show an interesting property which is very different from those in the $V_0$ case: for some quantum number $n$, the critical value of $m_{ex}$ in the $l=1$ case is larger than that in the $l=0$ case. The results in Fig.~\ref{Fig:alpha-on-mex-V1-monopole} also show a property different from the $V_0$ case: the boundaries of the regions are no longer straight. This indicates that the critical values of $m_{ex}$ and $\alpha$ do not satisfy a linear relation akin to Eq.~(\ref{equation:critical-screen-relation}) for finite $\Lambda$. Additionally, when $m_{ex}$ is fixed, the corresponding critical value of $\alpha$ decreases as the cutoff $\Lambda$ increases. This suggests that a smaller cutoff $\Lambda$ has a more significant impact on the existence of bound states.

\textcolor{black}{The results in Fig.~\ref{Fig:alpha-on-Lambda-V1-monopole}, with $m_{ex}=0.1$ and $\mu=1$, show that the critical values of $\alpha$ for the ground state depend only weakly on $\Lambda$. This suggests that varying $\Lambda$ does not substantially alter the number of bound states in the large-$\Lambda$ regime, especially for the ground state. To illustrate this explicitly, we give a short numerical table, Tab.~\ref{Tab:alpha-on-Lamda-V1-monopole}, where the trend can be read off directly.}
\begin{table}[htbp]
\centering
\begin{tabular}{c c c c c}
\hline
{$\Lambda$} & {$\alpha(l=0,n=1)$} & {$\alpha(l=0,n=2)$} & {$\alpha(l=0,n=3)$} & {$\alpha(l=0,n=4)$} \\
\hline
0.5 & 0.096 & 0.482 &1.173  &2.171 \\
1.0 & 0.088 & 0.395 &0.954  &1.760 \\
1.5 & 0.086 & 0.365 &0.874  &1.610 \\
2.0 & 0.085 & 0.351 &0.832  &1.530 \\
2.5 & 0.085 & 0.343 & 0.806 & 1.479 \\
3.0 & 0.085 & 0.338 & 0.788 & 1.443 \\
3.5 & 0.084 & 0.335 & 0.775 & 1.417 \\
4.0 & 0.084 & 0.332 & 0.766 & 1.397 \\
\hline
\end{tabular}
\caption{The critical values of $\alpha$ as functions of $\Lambda$ in the $V_1(r)$ case for $l=0$, where $\mu$ is fixed as $1$ and $m_{ex}$ is fixed as $0.1$.}
\label{Tab:alpha-on-Lamda-V1-monopole}
\end{table}

\subsection{Results for potential $V_2(r)$}\label{Sec:n4}
\textcolor{black}{As in the previous subsection, we use the dipole form factor with the same cutoff $\Lambda = 1$. Tab.~\ref{Tab:critical-mex-V2-dipole} shows that the critical values of $m_{ex}$ are smaller than those in Tab.~\ref{Tab:critical-mex-V1-monopole}. This is expected, because for a given nominal cutoff the strength and range of the effective interaction depend on whether a monopole or dipole form factor is adopted.}

\begin{figure}[htbp]
\centering
\includegraphics[bb=25 10 400 250, clip,scale=1.5]{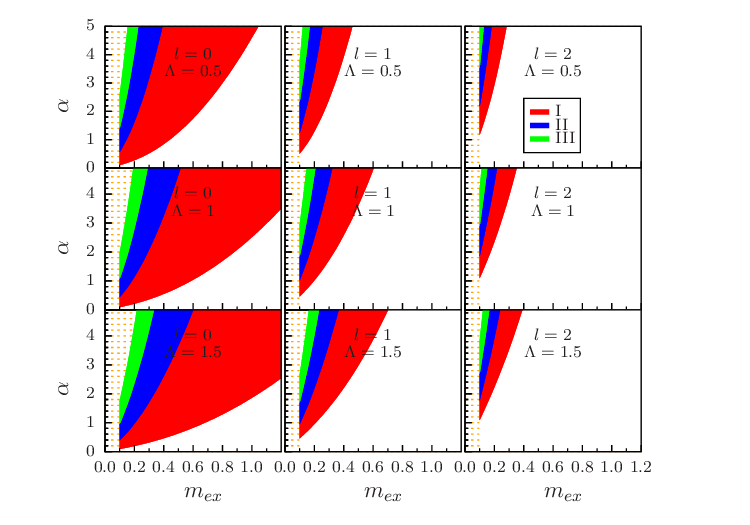}
\caption{Critical boundaries for the number of bound states as functions of $\alpha$ and $m_{ex}$ for specific angular momentum $l$ and cutoff $\Lambda$ in the $V_2(r)$ case, where $\mu$ is fixed as $1$. The panels, ordered from left to right, illustrate the cases for $l=0$, $l=1$, and $l=2$. The red (\RM1), blue (\RM2), and green (\RM3) regions signify the existence of one, two, and three bound states, respectively.}
\label{Fig:alpha-on-mex-V2-dipole}
\end{figure}

\begin{figure}[htbp]
\centering
\includegraphics[bb=25 140 400 250, clip,scale=1.5]{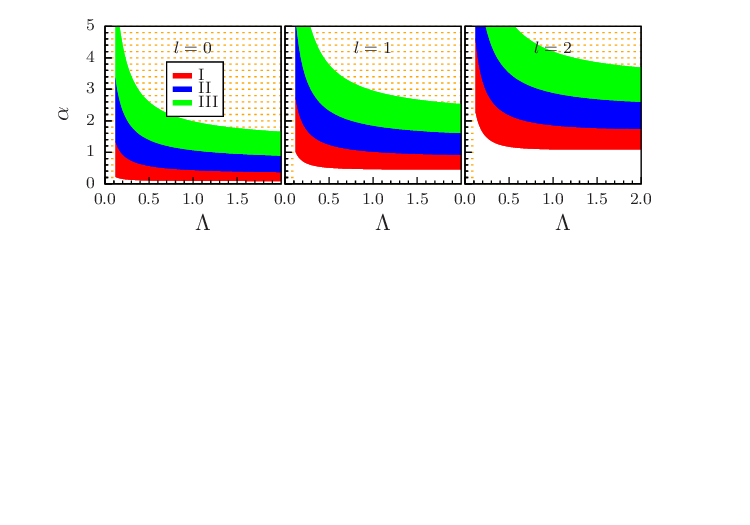}
\caption{Critical boundaries for the number of bound states as functions of $\alpha$ and $\Lambda$ for specific angular momentum $l$ in the $V_2(r)$ case, where $\mu$ is fixed as $1$ and $m_{ex}$ is fixed as $0.1$. The panels, ordered from left to right, illustrate the cases for $l=0$, $l=1$, and $l=2$. The red (\RM1), blue (\RM2), and green (\RM3) regions signify the existence of one, two, and three bound states, respectively.}
\label{Fig:alpha-on-Lambda-V2-dipole}
\end{figure}

\textcolor{black}{
Fig.~\ref{Fig:alpha-on-mex-V2-dipole} shows that the dependence of $\alpha$ on $m_{ex}$ is similar to that in Fig.~\ref{Fig:alpha-on-mex-V1-monopole}, while Fig.~\ref{Fig:alpha-on-Lambda-V2-dipole} closely parallels Fig.~\ref{Fig:alpha-on-Lambda-V1-monopole} for the dependence of $\alpha$ on $\Lambda$.
In addition, for the dipole form factor and the same $\Lambda$, a smaller $m_{ex}$ and a larger coupling $\alpha$ are required for the appearance of bound states. By readjusting $\Lambda$, similar critical values can be reproduced within our framework.
\renewcommand\tabcolsep{0.3cm}
\renewcommand{\arraystretch}{1.0}
\begin{table}[htbp]
\centering
\begin{tabular}{c|cc}\hline
\diagbox[height=0.3\baselineskip]{$n$}{$l$}& 0 & 1 \\\hline
 1 & 0.58558680838334368160717264884765735&--\\
 2& 0.18270070240342597051691912003895539&0.19689682790771643103189361708297792\\
 3& 0.095030571470798752697956881977611261&0.098383578056817713416914243429592524\\
 4& 0.058534282044609208097652179586460915&0.060272223040655954415361979835410004\\
 5& 0.039695180641185519311747813053258091&0.040805728711342126763446926499838344\\
 \hline
\end{tabular}
\caption{The critical values of $m_{ex}$ with $\mu=1$, $\alpha=1$, and $\Lambda=1$, categorized by quantum numbers $n$ and $l$ in the $V_2(r)$ case.}
\label{Tab:critical-mex-V2-dipole}
\end{table}
}

\subsection{Results for $D\bar{D}$ system}
In this subsection, we extend the above discussion to the $D\bar{D}$ system.
The kernels $K_{D\bar{D}}$ and potential $V_{D\bar{D}}$ introduced below follow Ref.~\cite{Wang:2021aql} for the $J=0$ $D\bar{D}$ channel; they should not be confused with the dipole form-factor potential denoted $V_2(r)$ in Sec.~\ref{Sec:n4}.
The energy spectrum of this system has been studied in Ref.~\cite{Wang:2021aql}, where the corresponding potential in the Schr\"{o}dinger equation for the $J=0$ state is given by:
\begin{eqnarray}
K_{D\bar{D}}(\boldsymbol{q})&=&ig^2\left[\frac{3}{4}\bar{K}(\boldsymbol{q},m_\rho)+
\frac{1}{4}\bar{K}(\boldsymbol{q},m_\omega)\right]+ig^2_{\sigma}\bar{K}(\boldsymbol{q},m_{\sigma}), \\
V_{D\bar{D}}(r) &=&-4\pi\alpha\left[\frac{3}{4}\bar{V}(r,m_\rho)+\frac{1}{4}\bar{V}(r,m_\omega)\right]-{4\pi}\alpha_{\sigma}
\bar{V}(r,m_{\sigma}),
\end{eqnarray}
where $\alpha\equiv \frac{g^2}{4\pi},\alpha_{\sigma}\equiv \frac{g_{\sigma}^2}{4\pi}$ and
\begin{eqnarray}
\bar{K}(\boldsymbol{q},m_{ex})&=&\frac{1}{\boldsymbol{q}^2+m_{ex}^2}
\left(\frac{\Lambda^2-m_{ex}^2}{\Lambda^2+\boldsymbol{q}^2}\right)^2, \\
\bar{V}(r,m_{ex}) &=&\frac{1}{8\pi r} \Big[2e^{-m_{ex}r}-2e^{-\Lambda r}+ e^{-\Lambda r}\frac{(m_{ex}^2-\Lambda^2)r}{\Lambda}  \Big].\label{equation:potential-DDbar}
\end{eqnarray}
We note that the form factor used here is not the dipole form, but rather the specific form adopted in Ref.~\cite{Wang:2021aql}.

In Ref.~\cite{Wang:2021aql}, the coupling constants are taken as $g_{\sigma}=-0.76$ and $g=5.247$ (or $\alpha=2.19$ and $\alpha_{\sigma}=0.046$). Since $\alpha_{\sigma}$ is much smaller than $\alpha$, we fix its value in the following discussion to study the effects of $\alpha$ and $\Lambda$. The masses for the exchanged bosons are taken as $m_\rho=0.776$~GeV, $m_\omega=0.783$~GeV, and $m_\sigma=0.6$~GeV. The masses for $D$ and $\bar{D}$ are taken as $m_D=m_{\bar{D}}=1.867$ GeV.

\begin{figure}[htbp]
\centering
\includegraphics[bb=18 140 400 250, clip,scale=1.6]{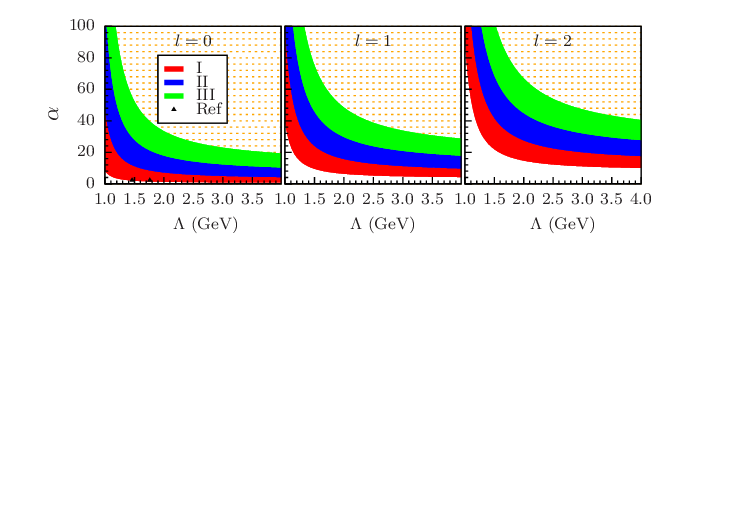}
\caption{Critical boundaries for the number of bound states as functions of $\alpha$ and $\Lambda$ for specific angular momentum $l$ in the $D\bar{D}$ system. The panels, ordered from left to right, illustrate the cases for $l=0$, $l=1$, and $l=2$. The red (\RM1), blue (\RM2), and green (\RM3) regions signify the existence of one, two, and three bound states, respectively.}
\label{Fig:alpha-on-Lambda-V-DDbar}
\end{figure}

In Fig.~\ref{Fig:alpha-on-Lambda-V-DDbar}, we show the critical boundaries for the number of bound states as functions of $\alpha$ and $\Lambda$, using the same notations as in Fig.~\ref{Fig:bound-states-alpha-mex-V0-monopole}. For comparison, the parameters used in Ref.~\cite{Wang:2021aql} are also presented by the two black triangles, which correspond to $\alpha=2.191$ with $\Lambda=1.46$~GeV and $\Lambda=1.76$~GeV.

\textcolor{black}{The results in Fig.~\ref{Fig:alpha-on-Lambda-V-DDbar} suggest that the critical value of $\alpha$ for the ground state depends weakly on $\Lambda$ when $\Lambda>2$~GeV. To show the detailed property, Tab.~\ref{Tab:alpha-on-Lamda-V-DDbar} gives the corresponding numbers, from which the variation of $\alpha$ with $\Lambda$ is seen to be much milder for $\Lambda\in[2,4]$~GeV than for $\Lambda\in[1,2]$~GeV.}

\begin{table}[htbp]
\centering

\begin{tabular}{c c c c c}
\hline
{$\Lambda$} & {$\alpha(l=0,n=1)$} & {$\alpha(l=0,n=2)$} & {$\alpha(l=0,n=3)$} & {$\alpha(l=0,n=4)$}\\
\hline
1.0   & 8.302 & 48.526 &121.416 &226.849 \\
1.5   & 1.927 & 11.241 &28.044  &52.311 \\
2.0   & 1.257 & 7.216  &17.920  &33.361 \\
2.5   & 1.025 & 5.758  &14.240  &26.476 \\
3.0   & 0.910 & 5.003  &12.336  &22.918 \\
3.5   & 0.843 & 4.538  &11.166  &20.736 \\
4.0   & 0.799 & 4.220  &10.371  & 19.252 \\
\hline
\end{tabular}
\caption{Critical values of $\alpha$ as functions of $\Lambda$ for the $l=0$ case in the $D\bar{D}$ system.}
\label{Tab:alpha-on-Lamda-V-DDbar}
\end{table}

\begin{figure}[htbp]
\centering
\includegraphics[scale=1.0]{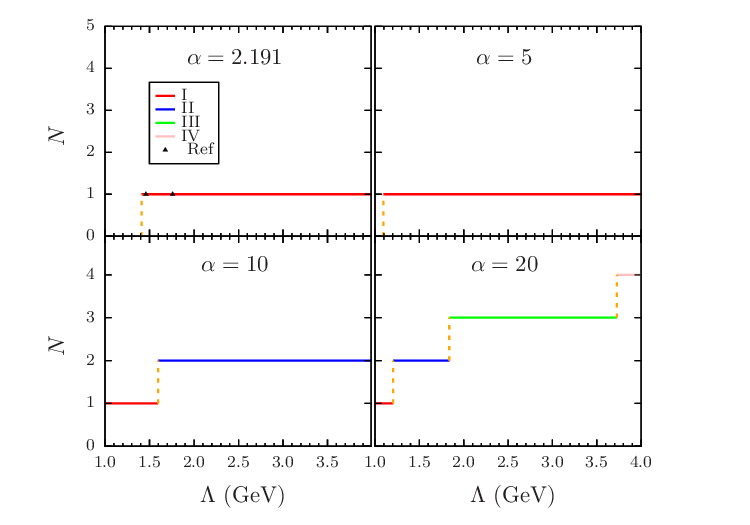}
\caption{The number of bound states $N$ as a function of $\Lambda$ for selected values of $\alpha$ at $l=0$ in the $D\bar{D}$ system.}
\label{Fig:N-on-Lambda-V-DDbar}
\end{figure}

\textcolor{black}{In Fig.~\ref{Fig:N-on-Lambda-V-DDbar}, we show the number of bound states $N$ as a function of $\Lambda$ for $\alpha=2.191$, 5, 10, and 20. For couplings typical of the present phenomenological context, $N$ remains equal to 1 when $\Lambda>1.4$~GeV. For $\alpha=5$, $N$ is also always 1 when $\Lambda>1.1$~GeV. For larger $\alpha$, $N$ increases with $\Lambda$.}

\begin{figure}[htbp]
\centering
\includegraphics[bb=10 130 400 250, clip,scale=1.55]{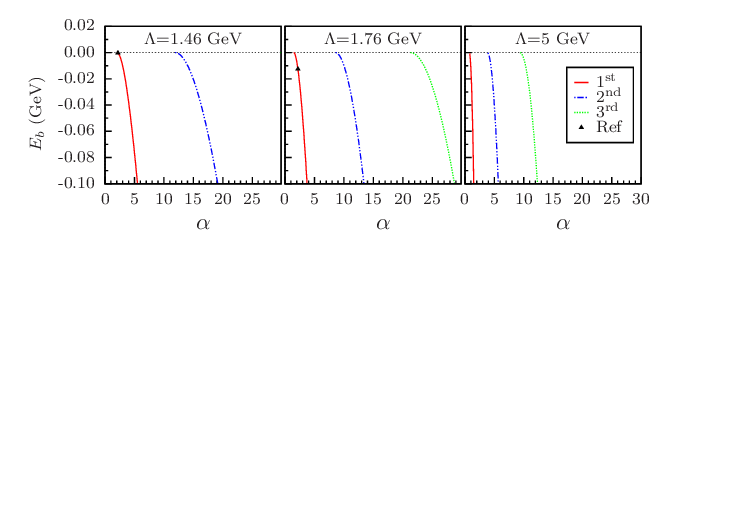}
\caption{Binding energy $E_b$ as a function of $\alpha$ with fixed $\Lambda$ for the $l=0$ case in the $D\bar{D}$ system. The panels, ordered from left to right, illustrate the cases for $\Lambda=1.46$~GeV, $\Lambda=1.76$~GeV, and $\Lambda=5$~GeV. The red, blue, and green lines signify the first, second, and third bound states, respectively. The two black triangles correspond to results using the parameters given in Ref.~\cite{Wang:2021aql}.}
\label{Fig:Eb-on-alpha-V-DDbar}
\end{figure}

In Fig.~\ref{Fig:Eb-on-alpha-V-DDbar}, we show the binding energy $E_b$ as a function of the coupling constant $\alpha$ with a fixed $\Lambda$ in the $l=0$ case, where the parameters used in Ref.~\cite{Wang:2021aql} are also presented by the two black triangles. The notations $1^{\text{st}}$, $2^{\text{nd}}$ and $3^{\text{rd}}$ refer to the first, second, and third bound states with $l=0$, respectively. The results show that $E_b$ is nearly linear in $\alpha$. As $\alpha$ increases, more bound states appear, which demonstrates that the magnitude of the coupling constant plays a critical role in determining the number of bound states at a fixed cutoff $\Lambda$.

\begin{figure}[htbp]
\centering
\includegraphics[bb=10 130 400 250, clip,scale=1.55]{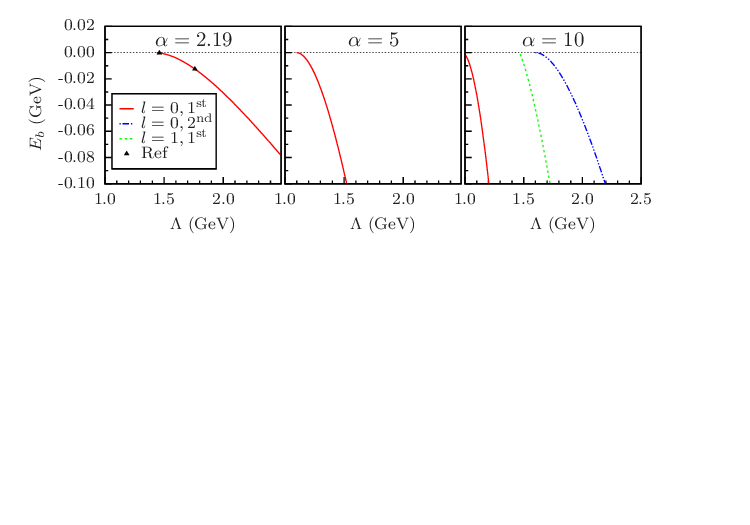}
\caption{Binding energy $E_b$ as a function of $\Lambda$ with fixed angular momentum $l$ and coupling constant $\alpha$ in the $D\bar{D}$ system. The panels, ordered from left to right, illustrate the cases for $\alpha=2.19$, $\alpha=5$, and $\alpha=10$. The red, blue, and green lines signify the first bound state with $l=0$, the second bound state with $l=0$, and the first bound state with $l=1$, respectively. The black triangle indicates the results using the parameters given in Ref.~\cite{Wang:2021aql}.}
\label{Fig:Eb-on-Lambda-V-DDbar}
\end{figure}

In Fig.~\ref{Fig:Eb-on-Lambda-V-DDbar}, we show the binding energy $E_b$ as a function of the cutoff $\Lambda$ with fixed coupling constant $\alpha$ and angular momentum $l$.  The results show that $E_b$ is nearly linear with respect to the cutoff $\Lambda$ when $\Lambda$ is large. When $\alpha$ is fixed as $2.19$ (the value used in Ref.~\cite{Wang:2021aql}), there is only one bound state.

\begin{figure}[htbp]
\centering
\includegraphics[scale=1.0]{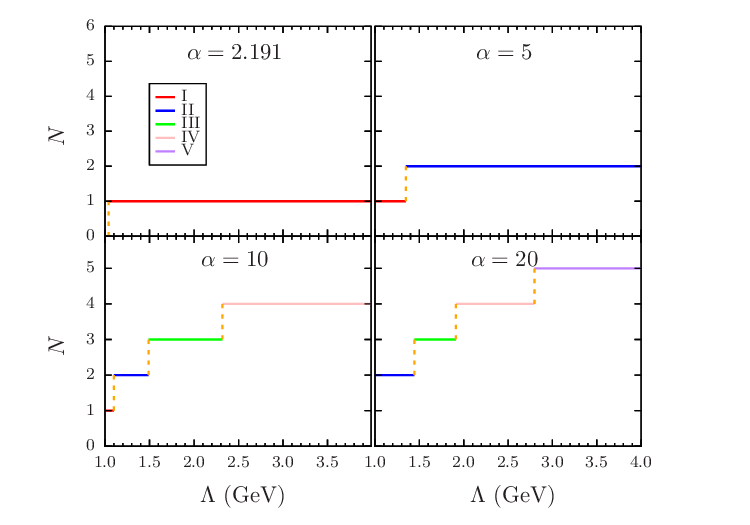}
\caption{The number of bound states $N$ as a function of $\Lambda$ for selected values of $\alpha$ at $l=0$ in the $B\bar{B}$ system, where the interaction parameters are taken from the $D\bar{D}$ analysis as a test.}
\label{Fig:N-on-Lambda-V-BBbar}
\end{figure}

\textcolor{black}{For comparison, we apply the same interaction parameters directly to the $B\bar{B}$ system. The corresponding results for the number of bound states $N$ are shown in Fig.~\ref{Fig:N-on-Lambda-V-BBbar}, which exhibit trends similar to those in the $D\bar{D}$ case.}

Combining all these results, we can conclude that the existence and multiplicity of bound states are strongly correlated with the coupling constant $\alpha$ especially when $\alpha$ is small, but only weakly dependent on the cutoff $\Lambda$ in a large region.

\section{Summary}\label{Sec:Summary}

In this study, we investigate the critical behavior for bound states in the Yukawa potential and a modified version, which are widely used in the OBE model for hadronic molecular states. The analysis is based on a highly accurate numerical method. Our numerical result for the critical mass $m_{ex}$ of the ground state is consistent with those reported in the literature. Furthermore, our results for the $l=1$ case are significantly more precise than those previously reported. \textcolor{black}{For the $D\bar{D}$ hadronic molecular system, our results indicate that the number of hadronic molecular states is only weakly sensitive to the cutoff $\Lambda$ over a broad window (for example, $\alpha\le5$ and $\Lambda>1.5~\mathrm{GeV}$), while binding energies may still depend on $\Lambda$, as discussed above.} This behavior suggests an interesting global regularity for hadronic molecular states. Finally, this highly accurate numerical method can also be straightforwardly applied to higher $l$ cases or other systems.

\section*{Acknowledgements}

H.~Q.~Zhou thanks Zhi-Hui Guo for helpful discussions. This work was funded by the National Natural Science Foundation of China (NSFC) under Grants Nos.~12075058 and 12150013 and supported by the Postgraduate Research
\& Practice Innovation Program of Jiangsu Province under Grants
No.~KYCX25\_0420.

\bibliography{ref}

\end{document}